\documentclass[conference]{IEEEtran}
\IEEEoverridecommandlockouts
\usepackage{cite}
\usepackage{amsmath,amssymb,amsfonts}
\usepackage{algorithmic}
\usepackage{graphicx}
\usepackage{textcomp}
\usepackage{xcolor}
\usepackage{float}
\usepackage{enumitem}
\usepackage{hyperref} 
\def\BibTeX{{\rm B\kern-.05em{\sc i\kern-.025em b}\kern-.08em
    T\kern-.1667em\lower.7ex\hbox{E}\kern-.125emX}}

\makeatletter
 \let\old@ps@headings\ps@headings
 \let\old@ps@IEEEtitlepagestyle\ps@IEEEtitlepagestyle
 \def\confheader#1{%
 \def\ps@IEEEtitlepagestyle{%
 \old@ps@IEEEtitlepagestyle%
 \def\@oddhead{\strut\hfill#1\hfill\strut}%
 \def\@evenhead{\strut\hfill#1\hfill\strut}%
 }%
 \ps@headings%
 }
 \makeatother

\confheader{%
 2024 International Conference on Integrated Intelligence and Communication Systems (ICIICS)
 }

 \usepackage[pscoord]{eso-pic}
\newcommand{\placetextbox}[3]{
 \setbox0=\hbox{#3}
 \AddToShipoutPictureFG*{ \put(\LenToUnit{#1\paperwidth},\LenToUnit{#2\paperheight}){\vtop{{\null}\makebox[0pt][c]{#3}}}
 }
 }
 \placetextbox{.23}{0.055}{\small{979-8-3315-0496-0/24/\$31.00~\copyright 2024 IEEE}}

\usepackage{cite}
\usepackage{amsmath,amssymb,amsfonts}
\usepackage{algorithmic}
\usepackage{graphicx}
\usepackage{textcomp}
\def\BibTeX{{\rm B\kern-.05em{\sc i\kern-.025em b}\kern-.08em
    T\kern-.1667em\lower.7ex\hbox{E}\kern-.125emX}}

\begin{document}
    \makeatletter
    \newcommand{\linebreakand}{%
      \end{@IEEEauthorhalign}
      \hfill\mbox{}\par
      \mbox{}\hfill\begin{@IEEEauthorhalign}
    }
    \makeatother

\title{A Temporal Convolutional Network-based Approach for Network Intrusion Detection}

\author{
  \IEEEauthorblockN{Rukmini Nazre}
  \IEEEauthorblockA{\textit{Department of CSE,} \\
    \textit{School of Computational Sciences}\\
    \textit{COEP Technological University}\\
    nazrerukmini@gmail.com}
  \and
  \IEEEauthorblockN{Rujuta Budke}
  \IEEEauthorblockA{\textit{Department of CSE,} \\
    \textit{School of Computational Sciences}\\
    \textit{COEP Technological University}\\
    rujutabudke@gmail.com}
  \and
  \IEEEauthorblockN{Omkar Oak}
  \IEEEauthorblockA{\textit{Department of CSE,} \\
    \textit{School of Computational Sciences}\\
    \textit{COEP Technological University}\\
    omkarsoak@gmail.com}
  \linebreakand 
  \IEEEauthorblockN{Suraj Sawant}
  \IEEEauthorblockA{\textit{Department of CSE,} \\
    \textit{School of Computational Sciences}\\
    \textit{COEP Technological University}\\
    sts.comp@coeptech.ac.in}
  \and
  \IEEEauthorblockN{Amit Joshi}
  \IEEEauthorblockA{\textit{Department of CSE,} \\
    \textit{School of Computational Sciences}\\
    \textit{COEP Technological University}\\
    adj.comp@coeptech.ac.in}
}

\maketitle

\begin{abstract}
 Network intrusion detection is critical for securing modern networks, yet the complexity of network traffic poses significant challenges to traditional methods. This study proposes a Temporal Convolutional Network(TCN) model featuring a residual block architecture with dilated convolutions to capture dependencies in network traffic data while ensuring training stability. The TCN’s ability to process sequences in parallel enables faster, more accurate sequence modeling than Recurrent Neural Networks. Evaluated on the Edge-IIoTset dataset, which includes 15 classes with normal traffic and 14 cyberattack types, the proposed model achieved an accuracy of 96.72\% and a loss of 0.0688, outperforming 1D CNN, CNN-LSTM, CNN-GRU, CNN-BiLSTM, and CNN-GRU-LSTM models. A class-wise classification report, encompassing metrics such as recall, precision, accuracy, and F1-score, demonstrated the TCN model’s superior performance across varied attack categories, including Malware, Injection, and DDoS. These results underscore the model’s potential in addressing the complexities of network intrusion detection effectively.
\end{abstract}


\section{Introduction}

The rapid growth of IoT devices and edge computing has transformed network topologies, creating complex systems for data collection and processing near the source to improve smart city applications' efficiency. This shift to edge computing distributes computational resources closer to data sources but also brings significant security risks. The expanded attack surface of IoT networks makes them prime targets for cyberattacks, highlighting the need for strong intrusion detection systems (IDS) to protect sensitive information \cite{1}. These networks face an array of sophisticated cyber threats, including malware intrusions, Distributed Denial of Service (DDoS) attacks, and Man-In-The-Middle (MITM) exploits \cite{2}.

Previous research has extensively explored various approaches to address these security challenges. Classical Machine Learning (ML) algorithms, such as k-Nearest Neighbors (KNN), Support Vector Machines (SVM), and Random Forests (RF), have been implemented for intrusion detection \cite{3,4,5}. However, these traditional methods face several limitations in terms of their adaptability to evolving attack patterns, insufficient capability to handle heterogeneous traffic patterns, reduced effectiveness in processing temporal dependencies in network traffic, and challenges in real-time detection and scalability. Motivated by these challenges, this research aims to develop a more robust and efficient intrusion detection system specifically designed for edge computing environments. Our primary objectives include designing a novel IDS framework utilizing Temporal Convolution Networks (TCNs), evaluating their effectiveness in capturing sequential patterns within network traffic, comparing their performance against traditional CNN architectures, and validating the system's effectiveness using the comprehensive Edge-IIoTset dataset \cite{6}.\\

This work makes several significant contributions to the field of network security in edge computing environments. First, we develop a novel TCN-based architecture specifically optimized for network intrusion detection. Second, we provide a comprehensive comparative analysis of the proposed TCN model against conventional CNN architectures. Third, we conduct empirical validation using the Edge-IIoTset dataset, demonstrating superior detection accuracy and reduced false positive rates. Thus, we present a scalable framework that can be deployed across diverse edge computing scenarios.\\ 
The remainder of this paper is organized as follows. Section \ref{sec:related_work} presents a comprehensive review of related work in network intrusion detection. Section \ref{sec:methodology} details the proposed TCN-based methodology. Section \ref{sec:results} presents the experimental setup, evaluation measures, results and comparative analysis. Finally, Section \ref{sec:conclusion} concludes the paper and discusses future research directions.\\

\section{Related Works}\label{sec:related_work}
Traditional ML models are not used in Network Intrusion Detection as they fail to capture intricate patterns from complex network traffic. So, more advanced Deep Learning (DL) approaches have been developed, demonstrating significantly improved performance. CNNs are frequently used for network intrusion detection \cite{7}. By using convolutional layers, 1D CNN models are able to capture patterns in network traffic, such as the frequency of certain types of packets or anomalies. However, CNNs lack the ability to capture temporal dependencies, that are critical for analyzing network traffic sequences. To address this, hybrid models that combine CNN with recurrent networks including Long Short-Term Memory (LSTM) and Gated Recurrent Units (GRU), have been explored. GRUs and LSTMs  are designed to handle time-series data by maintaining a memory of earlier inputs, making them effective for sequential data analysis\cite{8,9}. The use of Bidirectional LSTM (BiLSTM) further enhances temporal modeling by considering the context from both past and future inputs \cite{10}. Hybrid models like CNN-BiLSTM and CNN-GRU have demonstrated strong performance in various anomaly detection tasks but come with higher computational complexity and longer training times, especially on large datasets\cite{11}. Most of these models focus on specific attack types or simplified network environments, leaving room for more comprehensive solutions. \\
Recent advancements in DL have introduced TCNs\cite{12}, which offer an alternative to recurrent models. TCNs use one dimensional convolutional layers with dilated convolutions that help capture dependencies over a long range in data, enabling them to process sequential information in a parallelized manner. This architecture mitigates the common issues in recurrent networks, such as vanishing gradients and slow training times\cite{13}. Many studies have explored the usefulness of TCNs for time series prediction tasks, but their application in network intrusion detection still remains relatively new. The ability of TCNs to handle sequential data without relying on recurrence makes them particularly suitable for processing network traffic data, which can exhibit temporal patterns of normal and abnormal behavior. Their use provides an opportunity to enhance detection accuracy while maintaining faster processing times compared to traditional recurrent networks. Existing work on IDS in IoT environments has been centered primarily around specific attacks like DDoS \cite{14} using datasets such as IoT-23\cite{15} and CIC-IDS2018\cite{16}. These datasets, while useful, do not capture the comprehensive nature of IoT threats. The Edge-IIoTset dataset\cite{6} is a comprehensive benchmark for evaluating the performance of various models on 14 different types of attacks. The dataset’s rich collection of various types of attacks, such as malware, injection, MITM and scanning, has made it a reliable resource for testing the performance of various DL models \cite{17,18}. Existing studies have shown that while CNN-LSTM and CNN-GRU are effective at detecting DDoS attacks, they struggle with more complex attack vectors such as MITM or malware due to their reliance on temporal sequence modeling\cite{19}. TCNs offer a promising solution for this task\cite{20, 21}. They have been adopted for use on other datasets such as the APA-DDoS Dataset\cite{22} and the Bot-IoT Dataset\cite{23}. They have also been used for Binary Classification on the Edge-IIoTset dataset\cite{24}. However, there is a lack of studies focusing on TCN-based frameworks for Multiclass Classification on the EdgeIIoTset dataset.
This study proposes a TCN based model for 15-class classification on the EdgeIIoTSet dataset, and compares its performance with 1D CNN, CNN-LSTM, CNN-GRU, CNN-BiLSTM and CNN-GRU-LSTM.\\

\section{Proposed Method}\label{sec:methodology}
This section focuses on model development and dataset preparation. It covers the implementation of CNN, hybrid CNN models, and the proposed TCN model followed by a detailed description of the dataset, data preprocessing techniques, and the classification of output attack types.
\subsection{Prediction models}

\subsubsection{One-directional CNN}
1D CNNs \cite{25} process sequential data by applying convolutional filters along the temporal dimension, efficiently extracting local features that are useful for sequential data applications like language modeling, speech recognition, or time series analysis. However, their limited receptive field restricts their ability to capture long-range dependencies in the sequence, making it difficult to identify relationships between distant timesteps.

\subsubsection{Hybrid CNN}
Hybrid models like CNN-GRU, CNN-LSTM, and CNN-BiLSTM enhance 1D CNNs by integrating recurrent layers to better capture temporal dependencies. 
CNN-GRU offers a balance of performance and efficiency, while CNN-LSTM is effective for long-term dependencies but with higher computational costs.
CNN-BiLSTM processes sequences bidirectionally, improving context understanding but increasing training time. Among these, CNN-LSTM is often preferred for its robust long-term dependency management, despite its complexity. While hybrid models improve upon traditional 1D CNNs, their sequential nature limits parallelization and increases overfitting risks.

\subsubsection{Proposed Model}
The proposed model, shown in Figure \ref{fig3}, employs TCN to overcome the limitations of hybrid models like CNN-LSTM. TCNs utilize causal convolutions to ensure predictions depend only on past data, preserving temporal order and enabling efficient parallel processing.
Dilated convolutions expand the receptive field without increasing network depth, allowing TCNs to effectively capture both short and long range dependencies. TCN architecture has residual connections, which improve gradient flow and mitigate vanishing gradient issues, stabilizing training and enhancing representation learning.

\begin{figure}[h!]
\centering
\includegraphics[scale=0.1]
{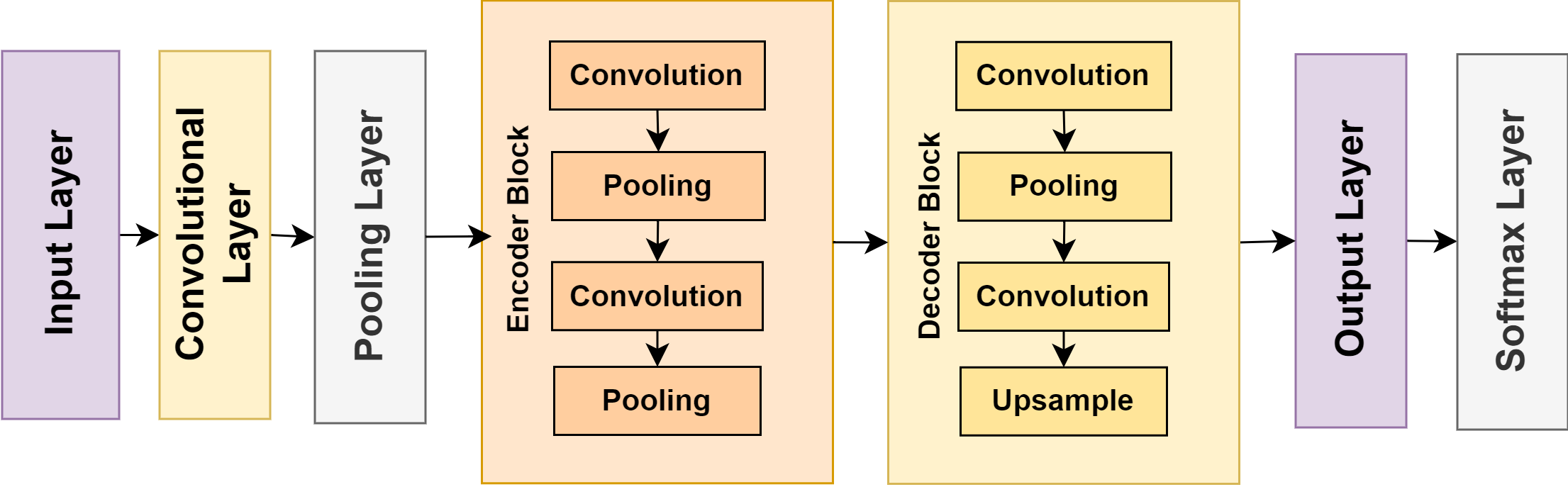}
    \caption{Architecture of the Proposed Model}
    \label{fig3}
\end{figure}

The proposed model consists of three stacked residual blocks with varying dilation rates to capture dependencies across different time scales. Following these blocks, the output is flattened, processed through a fully connected layer containing 128 neurons with dropout for regularization, and concludes with a softmax layer for classification. By integrating parallel computation, dilated convolutions, and residual connections, the proposed TCN model effectively manages long-term dependencies while minimizing computational costs.

\subsection{Dataset}\label{AA}
This study utilizes the publicly available Edge-IIoTset dataset\cite{6}, specifically the DNN-EdgeIIoT dataset, which consists of a CSV file containing 61 features along with two label columns— `Attack\_Label' and `Attack\_Type'. The dataset simulates 14 types of cyberattacks, classified into the following major subcategories: Information gathering, MITM, DoS/DDoS, Malware, and Injection attacks.
The dataset includes features enabling identification of network intrusion patterns such as TCP control flags, HTTP request types, ICMP sequence numbers, DNS query names, and MQTT topics.

\begin{table}[!ht]
    \centering
    \caption{List of All Label Classes}
    \scalebox{0.8}{
    \begin{tabular}{lllll}
        \hline
        \textbf{Attack Type}       & \textbf{Support Count} \\ \hline
        Normal                      & 349906                            \\
        DDoS\_UDP                   & 30392                             \\
        DDoS\_ICMP                  & 16985                             \\
        SQL\_injection              & 12706                             \\
        DDoS\_TCP                   & 12515                             \\
        Vulnerability\_scanner      & 12507                            \\
        Password                    & 12483                             \\
        DDoS\_HTTP                  & 12136                             \\
        Uploading                   & 9239                              \\
        Backdoor                    & 6007                              \\
        Port\_Scanning              & 4994                              \\
        XSS                         & 3767                              \\
        Ransomware                  & 2422                              \\
        Fingerprinting              & 213                               \\
        MITM                        & 90                                \\
        \hline
    \end{tabular}
    }
    \label{t1}
\end{table}

This study uses the dataset for multiclass classification, 15 classes specifically, which are shown in Table \ref{t1}. After preprocessing, the dataset contains 4,86,362 entries. A train, test and validation split of 70\%, 20\% and 10\% respectively is used. The training dataset has 3,40,452 values, test 97,273 and validation 48,637.

\subsection{Data Preprocessing}
\subsubsection{Encoding and Size Reduction}

\begin{table}[!ht]
    \centering
    \caption{List of Input Features}
    \scalebox{0.65}{
    \begin{tabular}{lll}
        \hline      
      \textbf{Feature}          & \textbf{Feature}          & \textbf{Feature}          \\ \hline
        1. tcp.ack                & 32. http1\_2              & 63. dns\_6                \\
        2. tcp.ack\_raw           & 33. http.response         & 64. mqtt3\_1              \\
        3. icmp.transmit\_timestamp & 34. dns\_0              & 65. mqtt2\_1              \\
        4. tcp.seq                & 35. mqtt3\_0              & 66. mbtcp.transmit\_request \\
        5. udp.stream             & 36. mqtt1\_0              & 67. mbtcp.transmit\_request\_in \\
        6. dns.qry.name           & 37. mqtt2\_0              & 68. dns\_4                \\
        7. icmp.checksum          & 38. tcp.connection.syn    & 69. dns\_6                \\
        8. icmp.seq\_le           & 39. http3\_5              & 70. http1\_7              \\
        9. tcp.dport              & 40. http1\_1              & 71. http3\_7              \\
        10. tcp.checksum          & 41. tcp.flags.ack         & 72. http3\_3              \\
        11. mqtt.hdrflags         & 42. http1\_4              & 73. http3\_10             \\
        12. tcp.len               & 43. http1\_encoded        & 74. dns\_9                \\
        13. http.content\_length  & 44. http2\_encoded        & 75. dns\_5                \\
        14. udp.time\_delta       & 45. http2\_2              & 76. mqtt1\_1              \\
        15. mqtt.len              & 46. mqtt.proto\_len       & 77. http3\_1              \\
        16. mqtt1.len             & 47. mqtt.confflags        & 78. http1\_8              \\
        17. tcp.flags             & 48. http2\_3              & 79. dns.retransmit\_request \\
        18. http1\_2              & 49. mqtt3\_encoded        & 80. mqtt1\_11             \\
        19. mqtt1\_1              & 50. http3\_2              & 81. dns\_8                \\
        20. http1\_0              & 51. tcp.connection.fin    & 82. mqtt3\_2              \\
        21. mqtt1\_encoded        & 52. http4\_4              & 83. http5\_1              \\
        22. mqtt3\_2              & 53. http4\_5              & 84. http5\_2              \\
        23. dns\_1                & 54. mqtt.conf.cleaness    & 85. mqtt1\_3              \\
        24. dns\_encoded          & 55. mqtt1\_2              & 86. http3\_2              \\
        25. dns\_3                & 56. mqtt3\_2              & 87. mqtt1\_4              \\
        26. mqtt3\_encoded        & 57. mqtt2\_2              & 88. http5\_1              \\
        27. mqtt2\_encoded        & 58. dns.retransmit        & 89. http3\_2              \\
        28. mqtt2\_encoded        & 59. mbtcp.trans\_id       & 90. dns\_7                \\
        29. mqtt.topic\_len       & 60. mbtcp.unit\_id        & 91. http3\_5              \\
        30. mqtt.msgtype          & 61. mbtcp.func\_id        & 92. mqtt1\_5              \\
        31. http1\_0              & 62. http3\_12             &                           \\ \hline
    \end{tabular}
    }
    \label{t2}
\end{table}

Label encoding is applied to categorical features related to HTTP requests, DNS queries, and MQTT protocol fields. Figure \ref{t2} contains all of the features used for classification. The features are one-hot encoded to prepare them for model training. Duplicate rows are identified and removed to maintain data integrity. Additionally, a hash-based method is used to find columns with identical content, and the duplicate columns are dropped. The Chi-Squared test is employed to identify and rank features based on their significance in relation to the target variable. The Chi-Squared test is a statistical test that evaluates the relationship between two categorical variables. It is chosen for feature selection because it can help determine which features are most strongly associated with the target variable - 'Attack\_type'. By ranking the features based on their Chi-Squared statistic, the most informative features can be identified and selected for model training. Certain columns irrelevant to the analysis, such as timestamps, IP addresses, as well as specific protocol data fields are removed. The distribution of different attack types is confirmed through an inspection of the attack type labels, and the redundant icmp.unused column is removed for its lack of meaningful contribution. To optimize the dataset for model training, a size reduction is implemented using stratified sampling. The dataset is reduced by a factor of 0.25, ensuring that the class distribution of the `Attack\_type' column remains consistent.

\subsubsection{Feature Scaling}
In this stage, feature scaling is performed using standardization to ensure that the input features have a standard deviation equal to one and a mean equal to zero. The StandardScaler is fitted to the training set, and subsequently applied to transform the test, validation, and training sets. This helps to mitigate issues related to differing scales of the features, thereby improving the performance and convergence speed of the models. By standardizing the data, this study ensures that each feature contributes equally to the distance calculations in algorithms that rely on distance metrics.

\subsection{Output Classes}
The dataset includes both normal and malicious network traffic, categorized into various attack types. Normal traffic corresponds to communication or data exchanges which are non-malicious and legitimate. DDoS attacks flood systems with excessive traffic, preventing normal operations. These include DDoS attacks targeting different network protocols such as ICMP, UDP, TCP and HTTP. 
SQL Injection allows attackers to manipulate databases through malicious queries, while Uploading involves uploading harmful files to compromise systems. Cross-Site Scripting (XSS) injects malicious scripts into web pages, leading to data theft or session hijacking. Vulnerability Scanners and Port Scanning detect system weaknesses, while Fingerprinting gathers system data for exploitation. 
Password Attacks focus on unauthorized access by cracking credentials, and Malware includes Backdoor attacks, which bypass authentication, and Ransomware, which encrypts files for ransom. Lastly, MITM attacks attempt to alter as well as intercept communications taking place amongst two entities. These attack types highlight key methods by which systems and networks can be compromised, essential for understanding cybersecurity threats.

\section{Results and Discussion}\label{sec:results}
This section discusses the experimental setup, evaluation measures and results, as well as a comparison with existing research.

\subsection{Experimental Setup}

The models are trained and tested in identical environments using Google Colab, equipped with an NVIDIA T4 GPU accelerator. The software requirements consist of Python 3.10 or a higher version, along with TensorFlow 2.17. The models are trained for 5 epochs using a learning rate of 0.001 and a batch size of 32. The Adam optimizer is used, and the loss function is Sparse Categorical Crossentropy.

\subsection{Evaluation Measures}

\subsubsection{Overall Accuracy}
Overall accuracy represents the percentage of correct predictions made by the model on the dataset, as shown in Equation \ref{eq:accuracy}.
\begin{equation}
\label{eq:accuracy}
\small
\text{Accuracy} = \frac{TP + TN}{TP + TN + FP + FN}
\end{equation}

Where \(TP\), \(TN\), \(FP\), and \(FN\) represent true positives, true negatives, false positives, and false negatives, respectively.

\subsubsection{Overall Loss - Sparse Categorical Cross Entropy}
This loss function measures the difference between actual class labels and predicted probabilities in a multiclass classification model that uses class indices instead of one-hot encoding, as defined in Equation \ref{eq:loss}.
\begin{equation}
\label{eq:loss}
\small
\text{Loss} = -\frac{1}{N} \sum_{i=1}^{N} \log(p_{y_i})
\end{equation}
where \(N\) is the number of samples, \(y_i\) is the true class label for sample \(i\), and \(p_{y_i}\) is the predicted probability of the true class.

\subsubsection{Classification Report}
The classification report provides metrics for each class, including:

i) Precision: The number of positive predictions that are correct divided by the number of positive predictions overall, indicating accuracy, as shown in Equation \ref{eq:precision}.
\begin{equation}
\label{eq:precision}
\small
\text{Precision} = \frac{TP}{TP + FP}
\end{equation}

ii) Recall: The number of positives predicted correctly divided by the number of actual positives, showing how well the model identifies positive cases, as defined in Equation \ref{eq:recall}.
\begin{equation}
\label{eq:recall}
\small
\text{Recall} = \frac{TP}{TP + FN}
\end{equation}

iii) F1 Score: A metric providing a balanced performance measure between recall and precision, as described in Equation \ref{eq:f1score}.
\begin{equation}
\label{eq:f1score}
\small
F1 = 2 \times \frac{\text{Precision} \times \text{Recall}}{\text{Precision} + \text{Recall}}
\end{equation}

iv) Support: The number of actual occurrences of each class, showing how classes are distributed.

\subsection{Results and Comparison}
As shown in Table \ref{results}, the proposed TCN model outperforms all other architectures in both accuracy and loss. The TCN achieves the highest test accuracy of 96.72\% and the lowest test loss of 0.0668, demonstrating its superior ability to capture complex patterns and dependencies in the data. In contrast, the 1D CNN has an accuracy of 96.18\%, the lowest among all evaluated models and the highest test loss of 0.0760, suggesting that it struggles to fully capture the intricate relationships within the dataset.

\begin{table}[htbp]
\centering
\caption{Performance of Different Models}

\begin{tabular}{lcc}
\hline
\textbf{Model}        & \textbf{Test Accuracy} & \textbf{Test Loss} \\ \hline
1D CNN                & 0.9618                 & 0.0760             \\
CNN GRU               & 0.9638                 & 0.0732             \\
CNN LSTM              & 0.9635                 & 0.0739             \\
CNN BiLSTM            & 0.9640                 & 0.0756             \\
CNN LSTM GRU          & 0.9640                 & 0.0733             \\
TCN                   & \textbf{0.9672}                 & \textbf{0.0668}             \\ \hline
\end{tabular}

\label{results}
\end{table}

Models incorporating recurrent layers such as CNN-GRU, CNN-LSTM, and CNN-LSTM-GRU improve accuracy and reduce loss compared to the standalone CNN. CNN-LSTM-GRU matches the performance of CNN-BiLSTM at 96.40\% accuracy. However, none of these models surpass the performance of the TCN. While existing research has utilized TCNs in various contexts, none have specifically addressed the challenge of 15-way multiclass classification on the Edge-IIoTset dataset\cite{24}. This study demonstrates significant advancements over existing research in network intrusion detection on this dataset \cite{6}. 

The classification reports for the models in Tables \ref{t3}, \ref{t4}, \ref{t5}, \ref{t6}, \ref{t7} and \ref{t8} and the confusion matrix of the proposed model in Figure \ref{fig10} show that all models perform well on the highly represented `Normal' class. However, their performance on classes with lower representation such as `SQL\_injection' and `XSS', is poor. The TCN architecture demonstrates robust performance across both high-frequency and low-frequency class distributions, particularly excelling in classes like `SQL injection' and `Uploading' with a weighted F1-score of 0.97. 

The proposed model outperforms existing models due to the following key improvements. By using stacked residual blocks with varying dilation rates, the model captures dependencies across different time scales more effectively. The integration of dilated convolutions allows the model to process both immediate and distant patterns. Additionally, the residual connections facilitate the training of deeper networks, preventing gradient problems. With the incorporation of parallel computation, the model significantly enhances training efficiency, while the fully connected layer and softmax classification ensure robust performance on large datasets.

\begin{table}[!ht]
    \centering
    \caption{1D CNN Classification Report}
    \scalebox{0.85}{
    \begin{tabular}{lllll}
        \hline
        & \textbf{Precision} & \textbf{Recall} & \textbf{F1-Score} & \textbf{Support} \\ \hline
        Normal                & 1              & 1             & 1          & 69832  \\
        MITM                  & 1              & 0.94          & 0.97       & 18     \\
        Uploading             & 0.86           & 0.64          & 0.73       & 1872   \\
        Ransomware            & 0.93           & 0.97          & 0.95       & 479    \\
        SQL\_injection        & 0.54           & 1             & 0.70       & 2500   \\
        DDoS\_HTTP            & 0.94           & 0.88          & 0.91       & 2507   \\
        DDoS\_TCP             & 0.98           & 0.92          & 0.95       & 2500   \\
        Password              & 0.93           & 0.33          & 0.49       & 2507   \\
        Port\_Scanning        & 0.83           & 1             & 0.91       & 1010   \\
        Vulnerability\_scanner & 0.98           & 0.91          & 0.95       & 2495   \\
        Backdoor              & 1              & 0.95          & 0.97       & 1221   \\
        XSS                   & 0.62           & 0.89          & 0.73       & 755    \\
        Fingerprinting        & 1              & 0.36          & 0.53       & 39     \\
        DDoS\_UDP             & 1              & 1             & 1          & 6005   \\
        DDoS\_ICMP            & 0.99           & 1             & 1          & 3533   \\  \hline       
        Accuracy              & ~              & ~             & 0.97       & 97273  \\
        Macro avg             & 0.91           & 0.85          & 0.85       & 97273  \\
        Weighted avg          & 0.98           & 0.97          & 0.97       & 97273  \\ \hline
    \end{tabular}
    }
    \label{t3}
\end{table}

\begin{table}[!ht]
    \centering
    \caption{CNN-GRU Classification Report}
    \scalebox{0.85}{
    \begin{tabular}{lllll}
    \hline
    & \textbf{Precision} & \textbf{Recall} & \textbf{F1-Score} & \textbf{Support} \\ \hline
    Normal                & 1.00           & 1.00          & 1.00       & 69832   \\
    MITM                  & 1.00           & 1.00          & 1.00       & 18      \\
    Uploading             & 0.88           & 0.52          & 0.66       & 1872    \\
    Ransomware            & 0.92           & 0.97          & 0.95       & 479     \\
    SQL\_injection        & 0.76           & 0.59          & 0.66       & 2500    \\
    DDoS\_HTTP            & 0.89           & 0.90          & 0.89       & 2507    \\
    DDoS\_TCP             & 0.98           & 0.92          & 0.95       & 2500    \\
    Password              & 0.50           & 0.77          & 0.61       & 2507    \\
    Port\_Scanning        & 0.84           & 1.00          & 0.91       & 1010    \\
    Vulnerability\_scanner & 0.96           & 0.92          & 0.94       & 2495    \\
    Backdoor              & 1.00           & 0.95          & 0.97       & 1221    \\
    XSS                   & 0.60           & 0.67          & 0.64       & 755     \\
    Fingerprinting        & 1.00           & 0.41          & 0.58       & 39      \\
    DDoS\_UDP             & 1.00           & 1.00          & 1.00       & 6005    \\
    DDoS\_ICMP            & 1.00           & 1.00          & 1.00       & 3533    \\  \hline       
    Accuracy              & ~              & ~             & 0.96       & 97273   \\
    Macro avg             & 0.89           & 0.84          & 0.85       & 97273   \\
    Weighted avg          & 0.97           & 0.96          & 0.96       & 97273   \\ \hline
\end{tabular}
}
\label{t4}
\end{table}

\begin{table}[!ht]
    \centering
    \caption{CNN-LSTM Classification Report}
    \scalebox{0.85}{
\begin{tabular}{lllll}
    \hline
    & \textbf{Precision} & \textbf{Recall} & \textbf{F1-Score} & \textbf{Support} \\ \hline
    Normal                & 1.00           & 1.00          & 1.00       & 69832   \\
    MITM                  & 1.00           & 1.00          & 1.00       & 18      \\
    Uploading             & 0.93           & 0.58          & 0.71       & 1872    \\
    Ransomware            & 0.79           & 0.90          & 0.84       & 479     \\
    SQL\_injection        & 0.70           & 0.66          & 0.68       & 2500    \\
    DDoS\_HTTP            & 0.79           & 0.96          & 0.87       & 2507    \\
    DDoS\_TCP             & 0.97           & 0.93          & 0.95       & 2500    \\
    Password              & 0.54           & 0.72          & 0.61       & 2507    \\
    Port\_Scanning        & 0.85           & 1.00          & 0.92       & 1010    \\
    Vulnerability\_scanner & 0.96           & 0.92          & 0.94       & 2495    \\
    Backdoor              & 0.98           & 0.88          & 0.93       & 1221    \\
    XSS                   & 0.78           & 0.33          & 0.46       & 755     \\
    Fingerprinting        & 1.00           & 0.18          & 0.30       & 39      \\
    DDoS\_UDP             & 1.00           & 1.00          & 1.00       & 6005    \\
    DDoS\_ICMP            & 0.99           & 1.00          & 1.00       & 3533    \\  \hline       
    Accuracy              & ~              & ~             & 0.96       & 97273   \\
    Macro avg             & 0.89           & 0.80          & 0.81       & 97273   \\
    Weighted avg          & 0.97           & 0.96          & 0.96       & 97273   \\ \hline
\end{tabular}
}
\label{t5}
\end{table}

\begin{table}[!ht]
    \centering
    \caption{CNN-BiLSTM Classification Report}
    \scalebox{0.85}{
\begin{tabular}{lllll}
    \hline
    & \textbf{Precision} & \textbf{Recall} & \textbf{F1-Score} & \textbf{Support} \\ \hline
    Normal                & 1.00           & 1.00          & 1.00       & 69832   \\
    MITM                  & 1.00           & 0.89          & 0.94       & 18      \\
    Uploading             & 0.81           & 0.55          & 0.66       & 1872    \\
    Ransomware            & 0.94           & 0.90          & 0.92       & 479     \\
    SQL\_injection        & 0.56           & 0.87          & 0.68       & 2500    \\
    DDoS\_HTTP            & 0.94           & 0.87          & 0.90       & 2507    \\
    DDoS\_TCP             & 0.97           & 0.94          & 0.95       & 2500    \\
    Password              & 0.65           & 0.44          & 0.53       & 2507    \\
    Port\_Scanning        & 0.87           & 0.96          & 0.91       & 1010    \\
    Vulnerability\_scanner & 0.96           & 0.92          & 0.94       & 2495    \\
    Backdoor              & 0.97           & 0.96          & 0.96       & 1221    \\
    XSS                   & 0.61           & 0.85          & 0.71       & 755     \\
    Fingerprinting        & 1.00           & 0.36          & 0.53       & 39      \\
    DDoS\_UDP             & 1.00           & 1.00          & 1.00       & 6005    \\
    DDoS\_ICMP            & 0.99           & 1.00          & 1.00       & 3533    \\  \hline       
    Accuracy              & ~              & ~             & 0.96       & 97273   \\
    Macro avg             & 0.88           & 0.83          & 0.84       & 97273   \\
    Weighted avg          & 0.97           & 0.96          & 0.96       & 97273   \\ \hline
\end{tabular}
}
\label{t6}
\end{table}

\begin{table}[!ht]
    \centering
    \caption{CNN-LSTM-GRU Classification Report}
    \scalebox{0.85}{
\begin{tabular}{lllll}
    \hline
    & \textbf{Precision} & \textbf{Recall} & \textbf{F1-Score} & \textbf{Support} \\ \hline
    Normal                & 1.00           & 1.00          & 1.00       & 69832   \\
    MITM                  & 1.00           & 0.89          & 0.94       & 18      \\
    Uploading             & 0.80           & 0.68          & 0.74       & 1872    \\
    Ransomware            & 0.82           & 0.78          & 0.80       & 479     \\
    SQL\_injection        & 0.54           & 0.98          & 0.70       & 2500    \\
    DDoS\_HTTP            & 0.91           & 0.90          & 0.90       & 2507    \\
    DDoS\_TCP             & 0.98           & 0.92          & 0.95       & 2500    \\
    Password              & 0.94           & 0.31          & 0.46       & 2507    \\
    Port\_Scanning        & 0.84           & 1.00          & 0.91       & 1010    \\
    Vulnerability\_scanner & 0.98           & 0.92          & 0.95       & 2495    \\
    Backdoor              & 0.93           & 0.90          & 0.91       & 1221    \\
    XSS                   & 0.60           & 0.73          & 0.66       & 755     \\
    Fingerprinting        & 1.00           & 0.36          & 0.53       & 39      \\
    DDoS\_UDP             & 1.00           & 1.00          & 1.00       & 6005    \\
    DDoS\_ICMP            & 0.99           & 1.00          & 1.00       & 3533    \\ \hline       
    Accuracy              & ~              & ~             & 0.96       & 97273   \\
    Macro avg             & 0.89           & 0.82          & 0.83       & 97273   \\
    Weighted avg          & 0.97           & 0.96          & 0.96       & 97273   \\ \hline
\end{tabular}
}
\label{t7}
\end{table}

\begin{table}[!ht]
    \centering
    \caption{Proposed TCN Model Classification Report}
    \scalebox{0.85}{
\begin{tabular}{lllll}
    \hline
    & \textbf{Precision} & \textbf{Recall} & \textbf{F1-Score} & \textbf{Support} \\ \hline
    Normal                & 1.00           & 1.00          & 1.00       & 69832   \\
    MITM                  & 1.00           & 1.00          & 1.00       & 18      \\
    Uploading             & 0.86           & 0.67          & 0.75       & 1872    \\
    Ransomware            & 0.92           & 0.97          & 0.95       & 479     \\
    SQL\_injection        & 0.55           & 1.00          & 0.71       & 2500    \\
    DDoS\_HTTP            & 0.93           & 0.89          & 0.91       & 2507    \\
    DDoS\_TCP             & 0.98           & 0.94          & 0.96       & 2500    \\
    Password              & 0.92           & 0.33          & 0.48       & 2507    \\
    Port\_Scanning        & 0.87           & 0.98          & 0.92       & 1010    \\
    Vulnerability\_scanner & 0.98           & 0.92          & 0.95       & 2495    \\
    Backdoor              & 1.00           & 0.95          & 0.97       & 1221    \\
    XSS                   & 0.61           & 0.83          & 0.71       & 755     \\
    Fingerprinting        & 1.00           & 0.36          & 0.53       & 39      \\
    DDoS\_UDP             & 1.00           & 1.00          & 1.00       & 6005    \\
    DDoS\_ICMP            & 0.99           & 1.00          & 1.00       & 3533    \\ \hline       
    Accuracy              & ~              & ~             & 0.97       & 97273   \\
    Macro avg             & 0.91           & 0.86          & 0.86       & 97273   \\
    Weighted avg          & 0.98           & 0.97          & 0.97       & 97273   \\ \hline
\end{tabular}}
\label{t8}
\end{table}

\begin{figure}[h!]
\centering
\includegraphics[scale=0.25]
{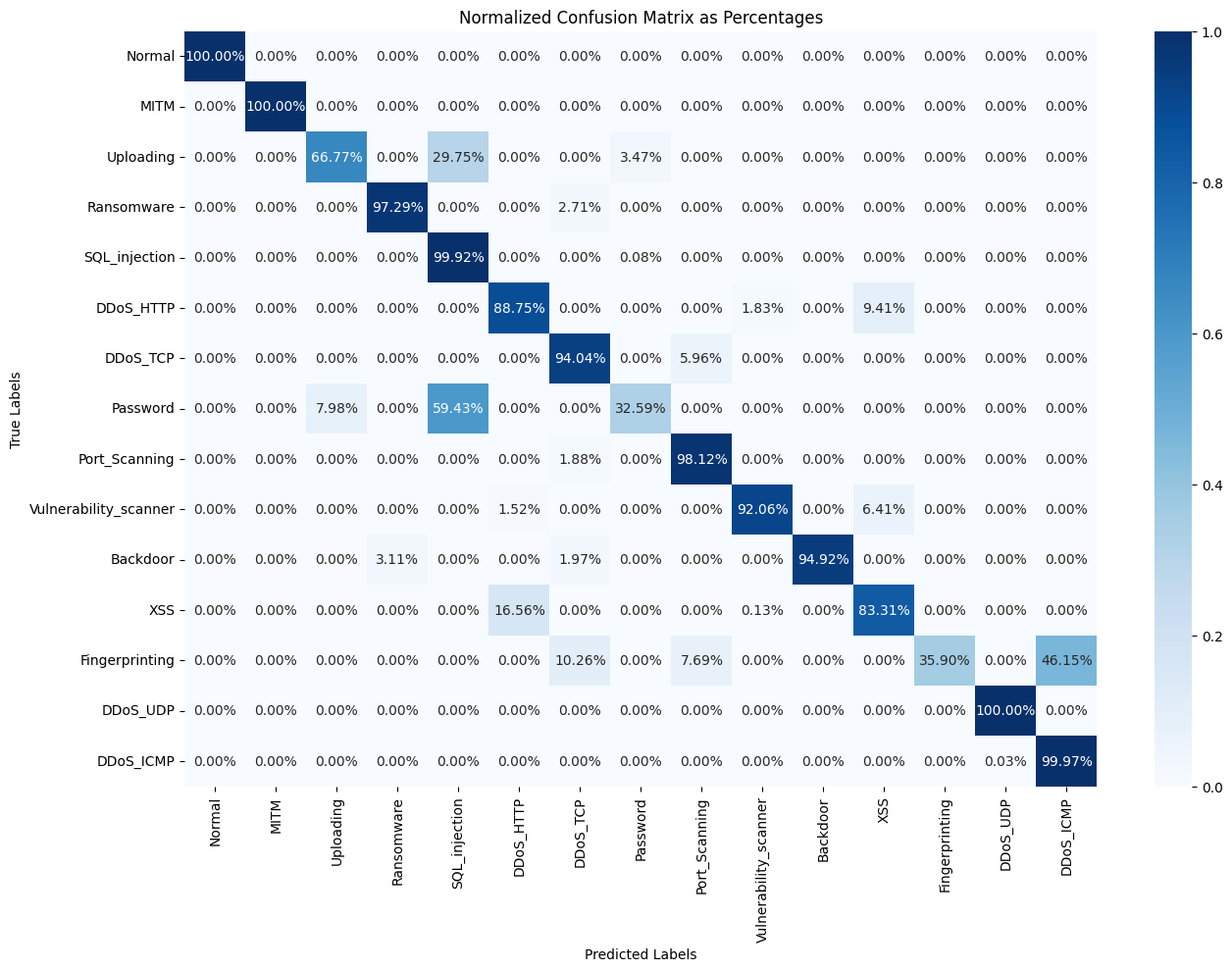}
    \caption{Confusion Matrix for the Proposed Model}
    \label{fig10}
\end{figure}

\section{Conclusion}\label{sec:conclusion}
This study proposes a novel TCN-based model for network intrusion detection, demonstrating strong performance on the Edge-IIoTset dataset. The TCN demonstrated superior performance, achieving the highest test accuracy of 96.72\% and the lowest test loss of 0.0668 compared to other existing models. The classification reports further highlight the TCN's effectiveness in accurately identifying a wide range of attack types, particularly excelling in detecting less frequent and more complex threats like SQL injection and uploading attacks. The enhanced performance of the TCN can be attributed to its ability to capture long-range dependencies and efficiently model temporal patterns within the network traffic data. Overall, the TCN-based model proves to be a robust and reliable solution for improving network security through accurate and efficient intrusion detection. Despite its strong performance, the proposed TCN-based model has certain limitations. It's evaluation is limited to the Edge-IIoTset dataset, which may restrict its generalizability to other datasets or real-time environments. The model's computational complexity could also challenge its deployment in resource-constrained settings. In future work, classification accuracy can be improved, especially for complex attacks. The scalability and performance can also be evaluated in dynamic conditions. Finally, the trust and usability of automated IDS in critical infrastructure can be increased, helping administrators better understand detection decisions.

\end{document}